\documentclass[pra, twocolumn, floatfix]{revtex4}
\usepackage{graphicx}
\usepackage{epstopdf}
\usepackage{amsmath, amsfonts, amssymb, bm}
\def\ifa{Institute of Applied Physics, Academiei str. 5, MD-2028 Chi\c{s}in\u{a}u, Moldova}
\begin{document}
\title{Dynamics of a quantum oscillator coupled with a three-level $\Lambda-$ type emitter}
\author{Alexandra \surname{M\^{i}rzac}}
\affiliation{\ifa}
\author{Mihai A. \surname{Macovei}}
\email{macovei@phys.asm.md}
\affiliation{\ifa}
\date{\today}
\begin{abstract}
We investigate the quantum dynamics of a quantum oscillator coupled with the most upper state of a three-level $\Lambda-$ type system. 
The two transitions of the three-level emitter, possessing orthogonal dipole moments, are coherently pumped with a single or two electromagnetic 
field  sources, respectively. We have found ranges for flexible lasing or cooling phenomena referring to the quantum oscillator's degrees 
of freedom. This is due to asymmetrical decay rates and quantum interference effects leading to population transfer among the relevant 
dressed states of the emitter's subsystem with which the quantum oscillator is coupled. As an appropriate system can be considered a 
nanomechanical resonator coupled with the most excited state of the three-level emitter fixed on it. Alternatively, if the upper state of 
the $\Lambda-$type system possesses a permanent dipole then it can couple with a cavity electromagnetic field mode which can be in 
the terahertz domain, for instance. In the latter case, we demonstrate an effective electromagnetic field source of terahertz photons.
\end{abstract}
\maketitle

\section{Introduction}
Lasing and cooling effects are among the most studied ones due to their enormous potential applications in the micro- or nano-world 
\cite{rew_art1,n_las,d_las,las_c,fon_l}. Presently, quantum technologies \cite{qt,qt1,qt2} require precise tools allowing a complete 
control of the quantum interaction between light and matter and, of course, the above mentioned phenomena occurring in a wide range 
of systems. Particularly, certain quantum systems offer additional control mechanisms 
via externally applied coherent light sources and, therefore, cooling phenomenon was successfully demonstrated in few-level atomic systems 
\cite{eit_theor,eit_cool,ek,plenio1}, for instance. On the other side, various optomechanical systems are intensively investigated recently because of their 
extreme sensitivity to ultra-weak perturbations \cite{rew_art2,rew_art3}. Thereby,  cooling or lasing in these systems are of fundamental interest as well 
\cite{las_c,xe,phon_las,cool_exp1,cool_exp2}. Furthermore, artificially created atomiclike systems such as quantum dots or quantum wells are also suitable 
for modern applications and exhibit an advantage with respect to engineering of their dipole moments, transition frequencies, etc. \cite{qwell,qdots,victor}. 
In these circumstances, ground-state cooling of a nanomechanical resonator with a triple quantum dot via quantum interference effects was demonstrated 
in \cite{gxl1}, see also \cite{gxl2,plenio2,plenio3}. Enhanced nanomechanical resonator's phonon emission via multiple excited quantum dots was demonstrated 
as well, in Ref.~\cite{vpm}. Moreover, among other applications of these systems or various optoelectronical schemes is the generation of electromagnetic field 
in the terahertz domain. The importance of the terahertz waves towards sensing, imaging, spectroscopy or data communications is highly recognized 
\cite{thz1,thz2,thz3}. In this context, quantum systems possessing permanent dipoles were shown to generate terahertz light \cite{tera1,tera2,terra,tera3,tera4}. 
Additionally, they exhibit bare-state population inversion as well as multiple spectral lines and squeezing \cite{mmk,gagik,paspalakis}.

Thus, there is an increased interest for novel quantum systems exhibiting lasing in a broad parameter range or cooling of micro- or nano-scale devices. 
From this point of view, here, we investigate a laser pumped $\Lambda-$type three-level system the upper state of which is being coupled with a quantum 
oscillator described by a quantized single-mode boson field. More specifically, as a quantum oscillator can serve a vibrational mode of a nanomechanical 
resonator containing the three-level emitter or, respectively, an electromagnetic cavity mode field if the upper state of the three-level sample, embedded 
in the cavity, possesses a permanent dipole. The frequency of the quantum oscillator is significant smaller than all other frequencies involved to describe 
the model, however, it is of the order of the generalized Rabi frequency characterizing the laser-pumped three-level qubit. In concordance to the 
dressed-state picture of the three-level system, we have identified two resonance conditions determining the oscillator's quantum dynamics, namely, when 
the quantum oscillator's frequency is close to the doubled generalized Rabi frequency or just to the generalized Rabi frequency, respectively. Correspondingly, 
we treat these two situations separately. We have found steady-state lasing or cooling regimes in both situations for the quantum oscillator's field mode, 
however, for asymmetrical spontaneous decay rates corresponding to each three-level qubit's transition. The mechanisms responsible for these effects are 
completely different for the two situations. In the case when the doubled generalized Rabi frequency is close to the oscillator's one, the model is somehow 
similar to a two-level system interacting with a quantized field mode where the spontaneous decay pumps both levels. On the other side, if the oscillator's 
frequency lies near resonance with the generalized Rabi frequency, then the sample is close to an equidistant three-level system where the single-mode 
quantum oscillator interacts with both qubit's transitions. The latter situation includes single- or two-quanta processes accompanied by quantum interference 
effects among the involved dressed-states leading to deeper cooling regimes and flexible ranges for lasing effects. This is different from other related 
schemes based on electromagnetically induced transparency processes  \cite{gxl1,gxl2,plenio2,plenio3}. In the case the model contains an electromagnetic 
cavity mode, which describes the quantum oscillator, then its frequency can be in the terahertz domain and, thus, we demonstrate an effective coherent 
electromagnetic field source of such photons. While lasing or cooling effects are available for two-level systems as well \cite{d_las,las_c,fon_l}, three-level 
ones may have an advantage in the sense that show improved results for the same parameters involved. This may help when there are only certain 
accessible parameter ranges. Furthermore, certain realistic novel systems are described by a three-level model.  For instance, as a concrete $\Lambda-$type 
system may be taken a laser-pumped color center emitter embedded on a vibrating membrane where strong coupling strengths can be achieved via vacuum 
dispersive forces \cite{plenio3}. Few coupled quantum dots are appropriate systems too \cite{gxl2,qpd}. Also, as alternative systems can be asymmetrical 
real or artificial few-level molecules possessing permanent dipoles, $d_{\alpha\alpha}\not =0$ \cite{tera1,tera2,terra,tera3,tera4,mmk,gagik,paspalakis,gong,qw}. 
If $d_{11} \gg \{d_{22},d_{33}\}$, then an electromagnetic  resonator mode can couple with the upper state of the $\Lambda-$type system via its permanent 
dipole. 
\begin{figure}[t]
\includegraphics[height =7cm]{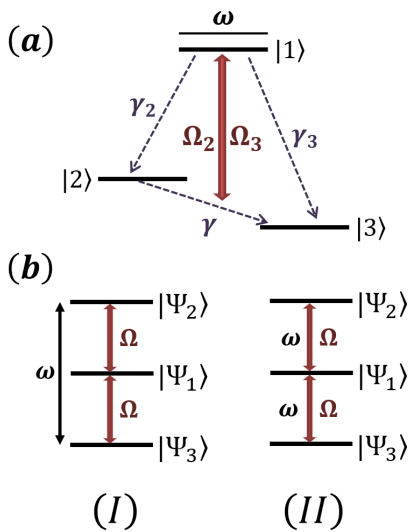}
\caption{\label{fig-1} (a) The schematic of the model: A laser pumped three-level $\Lambda-$type system the upper state of which, $|1\rangle$, 
is coupled with a quantum oscillator mode of frequency $\omega$. The oscillator can be described by a single mode of a nano-mechanical resonator 
containing the three-level emitter. Alternatively, if the upper state of the three-level system possesses a permanent dipole then it can couple 
with an electromagnetic cavity mode which can be in the terahertz ranges, for instance. Here, the pumping laser's frequencies are equal 
to the average transition frequency of the three-level emitter $(\omega_{12}+\omega_{13})/2$. $\Omega_{2}$ and $\Omega_{3}$ are the 
corresponding laser-qubit coupling strengths, i.e. the Rabi frequencies, whereas $\gamma's$ are the respective spontaneous decay rates.
(b) The semi-classical laser-qubit dressed-state picture where each bare-state level is dynamically split in three dressed-states 
$\{|\Psi_{2}\rangle,|\Psi_{1}\rangle,|\Psi_{3}\rangle\}$. Resonances occur at: (I) $\omega=2\Omega$ or (II) $\omega=\Omega$, respectively, 
where $\Omega$ is the generalized Rabi frequency.}
\end{figure}

The article is organized as follows. In Sec. II we describe the analytical approach and the system of interest, while in Sec. III we analyze the obtained results. 
The summary is given in Sec. IV.
\section{Theoretical framework}
The Hamiltonian describing a quantum oscillator of frequency $\omega$ coupled with a laser-pumped $\Lambda-$type three-level system, 
see Fig.~\ref{fig-1}(a), in a frame rotating at $(\omega_{12}+\omega_{13})/2$, is: 
\begin{eqnarray}
H&=& \hbar\omega b^{\dagger}b + \frac{\hbar\omega_{23}}{2}(S_{22} - S_{33})+\hbar g S_{11}(b+b^{\dagger}) \nonumber \\
&-& \hbar\sum_{\alpha \in \{2,3\}}\Omega_{\alpha}(S_{1\alpha} + S_{\alpha 1}).
\label{HM}
\end{eqnarray}
We have assumed here that as a pumping electromagnetic field source it can act a single laser of frequency $\omega_{L}$ pumping both arms of the 
emitter or, respectively, two lasers fields $\{\omega_{L1},\omega_{L2}\}$ each driving separately the two transitions of the $\Lambda-$ type sample 
possessing orthogonal transition dipoles. Additionally,  we have also considered that $\omega_{L1}=\omega_{L2} \equiv (\omega_{12}+\omega_{13})/2$, 
see Fig.~\ref{fig-1}(a). Here $\omega_{\alpha\beta}$ are the frequencies of $|\alpha\rangle \leftrightarrow |\beta\rangle$ three-level qubit's transitions, 
$\{\alpha,\beta \in 1,2,3\}$. The components entering in the Hamiltonian (\ref{HM}) have the usual meaning, namely, the first and the second terms 
describe the free energies of the quantum oscillator and the atomic subsystem, respectively, whereas the third one accounts for their mutual interaction 
via the most upper-state energy level with $g$ being the respective coupling strength. The last term represents the atom-laser interaction and 
$\{\Omega_{2},\Omega_{3}\}$ are the corresponding Rabi frequencies associated with a particular driven transition. Note that if the upper state of the 
investigated model contains a permanent dipole then the external coherent light sources interact with it as well. The corresponding Hamiltonian is:
$H_{pd}=\hbar S_{11}\sum_{i \in \{2,3\}}G_{i}\cos{(\omega_{Li}t)}$, where $G_{i}=d_{11}E_{i}/\hbar$ with $E_{i}$ being the lasers amplitudes. 
However, the Hamiltonian $H_{pd}$ can be considered as rapidly oscillating, because $\omega_{Li} \gg G_{i}$, and being further neglected. 
Thus, the Hamiltonian (\ref{HM}) and the analytical approach developed here allow to treat concomitantly both situations, namely, when either a 
nanomechanical resonator or an electromagnetic cavity is taken as a quantum oscillator. Finally, the three-level qubit's operators, 
$S_{\alpha\beta}=|\alpha\rangle\langle \beta|$, obey the commutation relation $[S_{\alpha\beta},S_{\beta'\alpha'}]$=$\delta_{\beta\beta'}S_{\alpha\alpha'}$ - $\delta_{\alpha'\alpha}S_{\beta'\beta}$ whereas those of the quantum oscillator's: $[b,b^{\dagger}]=1$ and $[b,b]=[b^{\dagger},b^{\dagger}]=0$, 
respectively.

In the Born-Markov approximations \cite{agarwal_2014,gardiner,wals}, the whole quantum dynamics of this complex model can be monitored via the 
following master equation:
\begin{eqnarray}
\dot \rho &+& \frac{i}{\hbar}[H,\rho]= - \sum_{\alpha \in \{2,3\}}\gamma_{\alpha}[S_{1\alpha},S_{\alpha 1}\rho] - \gamma[S_{23},S_{32}\rho]
\nonumber \\
&-&  \kappa(1+\bar n)[b^{\dagger},b\rho] - \kappa\bar n[b,b^{\dagger}\rho] + H.c..
\label{MEQ}
\end{eqnarray}
The right-hand side of Eq.~(\ref{MEQ}) describes the emitter's damping due to spontaneous emission as well as the quantum oscillator's damping effects 
with $\bar n=1/[\exp{(\hbar\omega/k_{B}T)}-1]$ being the mean oscillator's quanta number due to the environmental thermostat at temperature $T$. 
Here $k_{B}$ is the Boltzmann constant, $\gamma$'s are the corresponding decay rates of the three-level qubit, see Fig.~\ref{fig-1}(a), while $\kappa$ 
describes the quantum oscillator's leaking rate, respectively. The physics behind our model can be easier highlighted if we turn to the three-level qubit-laser 
dressed-state picture given by the transformation:
\begin{eqnarray}
|1\rangle &=& \sin{\theta}|\Psi_{1}\rangle - \frac{\cos{\theta}}{\sqrt{2}}\bigl(|\Psi_{2}\rangle + |\Psi_{3}\rangle\bigr),
\nonumber \\
|2\rangle &=& \frac{\cos{\theta}}{\sqrt{2}}|\Psi_{1}\rangle + \frac{1}{2}(1+\sin{\theta})|\Psi_{2}\rangle - \frac{1}{2}(1-\sin{\theta})|\Psi_{3}\rangle,
\nonumber \\
|3\rangle &=& -\frac{\cos{\theta}}{\sqrt{2}}|\Psi_{1}\rangle + \frac{1}{2}(1-\sin{\theta})|\Psi_{2}\rangle - \frac{1}{2}(1+\sin{\theta})|\Psi_{3}\rangle,
\nonumber \\ \label{DRS}
\end{eqnarray}
where $\sin{\theta}=\omega_{23}/(2\Omega)$ and $\cos{\theta}=\sqrt{2}\Omega_{0}/\Omega$ with 
$\Omega=\sqrt{2\Omega^{2}_{0}+(\omega_{23}/2)^{2}}$ being the generalized Rabi frequency whereas $\Omega_{2}=\Omega_{3}\equiv \Omega_{0}$. 
Applying the transformation (\ref{DRS}) to the Hamiltonian $(\ref{HM})$ one arrives at the corresponding Hamiltonian's expression in the dressed-state 
picture, i.e., $H=H_{0}+H_{d}+H_{1}+H_{2}$, where
\begin{eqnarray}
H_{0}&=&\hbar\omega b^{\dagger}b + \hbar\Omega R_{z},
\nonumber \\
H_{d}&=&\hbar g\bigl(\sin^{2}{\theta}R_{11} + \cos^{2}{\theta}(R_{22}+R_{33})/2\bigr)\bigl(b + b^{\dagger}\bigr), 
\nonumber \\
H_{1}&=&\hbar g\cos^{2}{\theta}\bigl(R_{32} + R_{23}\bigr)\bigl(b + b^{\dagger}\bigr)/2,
\nonumber \\
H_{2}&=&-\hbar g\frac{\sin{2\theta}}{2\sqrt{2}}\bigl(R_{21}+R_{13}+H.c.\bigr)\bigl(b + b^{\dagger}\bigr),
\label{DHM}
\end{eqnarray}
with $R_{z}=R_{22}-R_{33}$. Here the dressed-state three-level qubit's operators are: $R_{\alpha\beta}=|\Psi_{\alpha}\rangle\langle \Psi_{\beta}|$ and
obeying the same commutation relations as the old ones. In the interaction picture, characterized by the unitary operator 
\begin{eqnarray}
U(t)=\exp{(iH_{0}t/\hbar)}, \label{uo}
\end{eqnarray}
$H_{d}$ can be considered as a fast oscillating one and omitted from the dynamics, while the last two Hamiltonians transforms as:
\begin{eqnarray}
H_{1I}&=&\bar g\bigl(R_{23}e^{2i\Omega t}+H.c.\bigr)\bigl(b^{\dagger}e^{i\omega t}+ H.c.\bigr),
\nonumber \\
H_{2I}&=&-\tilde g\bigl((R_{21}+R_{13})e^{i\Omega t} + H.c.\bigr)\bigl(b^{\dagger}e^{i\omega t}+ H.c.\bigr),
\nonumber \\  \label{DHMI}
\end{eqnarray}
where 
\begin{eqnarray}
\bar g=\hbar g\cos^{2}{\theta}/2, \label{gbar} 
\end{eqnarray}
whereas 
\begin{eqnarray}
\tilde g = \hbar g\sin{2\theta}/(2\sqrt{2}). \label{gtilde}
\end{eqnarray}
Analyzing the above Hamiltonians one can observe that the quantum dynamics of our model is determined by two resonances (see Fig.~\ref{fig-1}b), 
namely, $(I)$ at 
\begin{eqnarray}
2\Omega = \omega,  \label{trez}
\end{eqnarray}
and $(II)$ at 
\begin{eqnarray}
\Omega=\omega. \label{rez}
\end{eqnarray}
Therefore, in what follows, we shall treat these two cases separately. Thus, the Hamiltonian for the first situation, $(I)$, will be
\begin{eqnarray}
H=\bar \delta b^{\dagger}b +  \bar g\bigl(R_{32}b^{\dagger} + b R_{23}\bigr),  \label{DH1}
\end{eqnarray}
while for the second case, $(II)$, is
\begin{eqnarray}
H=\tilde \delta b^{\dagger}b -  \tilde g\bigl((R_{12}+R_{31})b^{\dagger} +b(R_{21}+R_{13})\bigr),  \label{DH2}
\end{eqnarray}
where, respectively, $\bar \delta = \omega-2\Omega$ whereas $\tilde \delta = \omega - \Omega$.  Additionally, applying the dressed-state transformation 
(\ref{DRS}) to the corresponding damping part of the master equation (\ref{MEQ}), followed by the operation (\ref{uo}), one arrives at a  master equation, 
see Appendix A, which allows to obtain an exact system of equations describing the quantum dynamics of the examined system. Note that rapidly oscillating 
components in the above Hamiltonians, i.e. (\ref{DH1},\ref{DH2}), as well as in the final master equation (\ref{MEQA}) were dropped, meaning that 
$\Omega \gg \{g, \gamma,\gamma_{2},\gamma_{3}\}$. 

In what follows, we shall compare the two situations, i.e. $(I)$ and $(II)$, for the same parameters range and discuss the physics behind.
\section{Results and Discussions}
The equations of motion, for the first situation $(I)$, describing the oscillator's quantum dynamics (i.e., mean quanta number and its quantum statistics, 
qubit's populations etc.) can be obtained with the help of Eq.~(\ref{MEQA}):
\begin{eqnarray}
\dot P^{(0)}_{n} &=& i\bar g(P^{(5)}_{n} - P^{(3)}_{n}) - 2\kappa\bar n\bigl((n+1)P^{(0)}_{n} \nonumber \\
&-& nP^{(0)}_{n-1}\bigr) - 2\kappa(1+\bar n)\bigl(nP^{(0)}_{n} - (n+1) \nonumber \\
&\times& P^{(0)}_{n+1}\bigr), \nonumber  \\
\dot P^{(1)}_{n} &=& i\bar g(P^{(5)}_{n} - P^{(3)}_{n}) - 2\kappa\bar n\bigl((n+1)P^{(1)}_{n} \nonumber \\
&-&  nP^{(1)}_{n-1}\bigr) - 2\kappa(1+\bar n)\bigl(nP^{(1)}_{n} - (n+1) \nonumber \\
&\times& P^{(1)}_{n+1}\bigr) + \gamma^{(1)}_{0}P^{(0)}_{n} - \gamma^{(1)}_{1}P^{(1)}_{n}, \nonumber \\
\dot P^{(2)}_{n} &=& i\bar g(P^{(5)}_{n} + P^{(3)}_{n}) - 2\kappa\bar n\bigl((n+1)P^{(2)}_{n} \nonumber \\
&-& nP^{(2)}_{n-1}\bigr) - 2\kappa(1+\bar n)\bigl(nP^{(2)}_{n} - (n+1)  \nonumber \\
&\times&P^{(2)}_{n+1}\bigr) + \gamma^{(2)}_{0}P^{(0)}_{n} - \gamma^{(2)}_{1}P^{(1)}_{n} 
- \gamma^{(2)}_{2}P^{(2)}_{n}, \nonumber \\
\dot P^{(3)}_{n} &=& i\bar \delta P^{(4)}_{n} - i\bar gn(P^{(1)}_{n}-P^{(2)}_{n}-P^{(1)}_{n-1} \nonumber \\
&-&P^{(2)}_{n-1}) -\kappa(1+\bar n)\bigl((2n-1)P^{(3)}_{n} -2 (n+1) \nonumber \\
&\times&P^{(3)}_{n+1} + 2P^{(5)}_{n}\bigr) - \kappa\bar n\bigl((2n+1)P^{(3)}_{n}  \nonumber \\
&-& 2nP^{(3)}_{n-1}\bigr) - \gamma^{(3)}_{3}P^{(3)}_{n}, \nonumber 
\end{eqnarray}
\begin{eqnarray}
\dot P^{(4)}_{n} &=& i\bar \delta P^{(3)}_{n} -\kappa(1+\bar n)\bigl((2n-1)P^{(4)}_{n} + 2P^{(6)}_{n} \nonumber \\
&-& 2 (n+1)P^{(4)}_{n+1}\bigr) - \kappa\bar n\bigl((2n+1)P^{(4)}_{n} \nonumber \\
&-&2nP^{(4)}_{n-1}\bigr)- \gamma^{(4)}_{4}P^{(4)}_{n}, \nonumber \\
\dot P^{(5)}_{n} &=& i\bar \delta P^{(6)}_{n} + i\bar g(n+1)(P^{(1)}_{n} + P^{(2)}_{n} - P^{(1)}_{n+1} \nonumber \\
&+& P^{(2)}_{n+1}) - \kappa(1+\bar n)\bigl((2n+1)P^{(5)}_{n} \nonumber \\
&-&2 (n+1)P^{(5)}_{n+1}\bigr) - \kappa\bar n\bigl((2n+3)P^{(5)}_{n} \nonumber \\
&-& 2nP^{(5)}_{n-1} - 2P^{(3)}_{n}\bigr)- \gamma^{(5)}_{5}P^{(5)}_{n}, \nonumber \\
\dot P^{(6)}_{n} &=& i\bar \delta P^{(5)}_{n} -  \kappa\bar n\bigl((2n+3)P^{(6)}_{n} - 2nP^{(6)}_{n-1}  \nonumber \\
&-&2P^{(4)}_{n}\bigr) - \kappa(1+\bar n)\bigl((2n+1)P^{(6)}_{n} \nonumber \\
&-& 2(n+1)P^{(6)}_{n+1}\bigr)- \gamma^{(6)}_{6}P^{(6)}_{n}. \label{EQM1}
\end{eqnarray}
Here $\gamma^{(1)}_{0}=\bigl((\gamma^{(-)} + \gamma^{(+)})\sin^{2}{\theta} + \gamma\cos^{2}{\theta}(1+\sin^{2}{\theta})\bigr)/2$,
$\gamma^{(1)}_{1}=2\gamma^{(0)}_{0}+(\gamma^{(-)} + \gamma^{(+)})\sin^{2}{\theta}/2 + 3\gamma\cos^{2}{\theta}(1+\sin^{2}{\theta})/4$,
$\gamma^{(2)}_{0}=\bigl((\gamma^{(+)} - \gamma^{(-)})\sin^{2}{\theta} - 2\gamma\sin{\theta}\cos^{2}{\theta}\bigr)/2$,
$\gamma^{(2)}_{1}=2(\Gamma^{(-)} - \Gamma^{(+)}) + (\gamma^{(+)} - \gamma^{(-)})\sin^{2}{\theta}/2 -\gamma\sin{\theta}\cos^{2}{\theta}/2$,
$\gamma^{(2)}_{2}=2\bigl(\gamma^{(0)}_{0}+\Gamma^{(-)} + \Gamma^{(+)} + \gamma\cos^{2}{\theta}(1+\sin^{2}{\theta})/8\bigr)$ and 
$\gamma^{(3)}_{3} = (\gamma_{2}+\gamma_{3})\cos^{2}{\theta}/2 + 2\gamma^{(0)}_{0} + \Gamma^{(-)} + \Gamma^{(+)} + 
\gamma\cos^{2}{\theta}(1+\sin^{2}{\theta})/4$ with $\gamma^{(4)}_{4}=\gamma^{(5)}_{5}=\gamma^{(6)}_{6}=\gamma^{(3)}_{3}$. 
Further, $\gamma^{(\pm)}=\gamma_{2}(1 \pm \sin{\theta})^{2} + \gamma_{3}(1 \mp \sin{\theta})^{2}$, 
$\Gamma^{(\pm)}=\gamma^{(\pm)}\cos^{2}{\theta}/8 + \gamma(1 \mp \sin{\theta})^{4}/16$,
$\gamma^{(\pm)}_{0}=\pm\bigl(\gamma_{3}(1 \mp \sin{\theta}) - \gamma_{2}(1 \pm \sin{\theta})\bigr)\sin{\theta}\cos^{2}{\theta}/2$ 
and $\gamma^{(0)}_{0} =(\gamma_{2}+\gamma_{3})\cos^{4}{\theta}/4$. 
To arrive at the system of equations (\ref{EQM1}), first we obtained the corresponding equations for variables: $\rho^{(0)}=\rho_{11}+\rho_{22}+\rho_{33}$, 
$\rho^{(1)}=\rho_{22}+\rho_{33}$, $\rho^{(2)}=\rho_{22}-\rho_{33}$, $\rho^{(3)}=b^{\dagger}\rho_{23}-\rho_{32}b$, 
$\rho^{(4)}=b^{\dagger}\rho_{23} + \rho_{32}b$, $\rho^{(5)}=\rho_{23}b^{\dagger} - b\rho_{32}$, $\rho^{(6)}=\rho_{23}b^{\dagger} + b\rho_{32}$, 
where $\rho_{\alpha\beta}=\langle \alpha|\rho|\beta\rangle$, and then projecting on the Fock states $|n\rangle$, i.e., 
$P^{(i)}_{n}=\langle n|\rho^{(i)}|n\rangle$, $\{i \in 0 \cdots 6\}$ and $n \in \{0, \infty\}$, see also \cite{quang}. Thus, the analytical approach developed 
here allows us to obtain an exact system of equations describing the quantum dynamics of the composed system {\it laser pumped spontaneously damped 
qubit plus leaking phonon mode} within the rotating wave, Born-Markov and secular approximations, respectively, and to extract the variables of interest 
with the help of the traced density operator over the corresponding degrees of freedom.
\begin{figure}[t]
\includegraphics[width = 4.23cm]{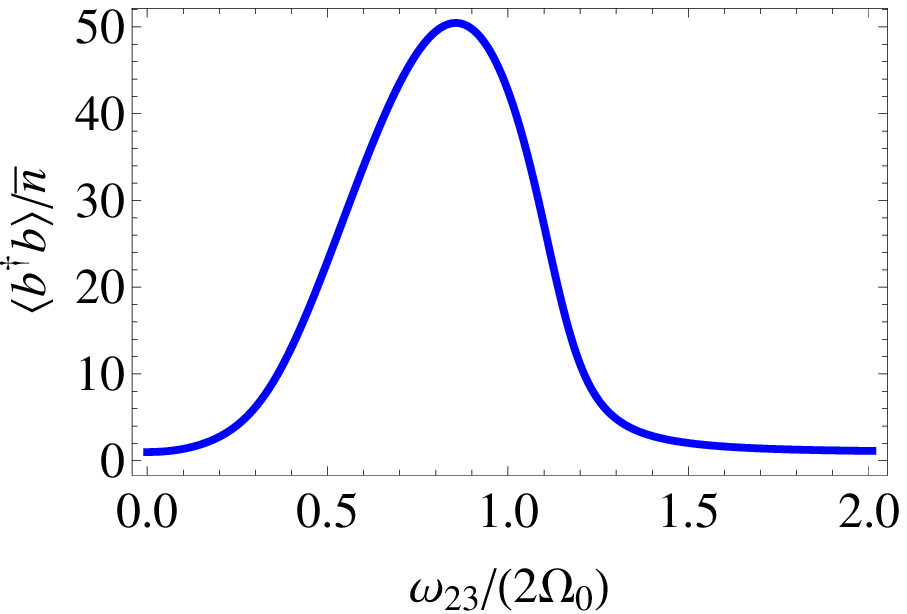}
\includegraphics[width = 4.23cm]{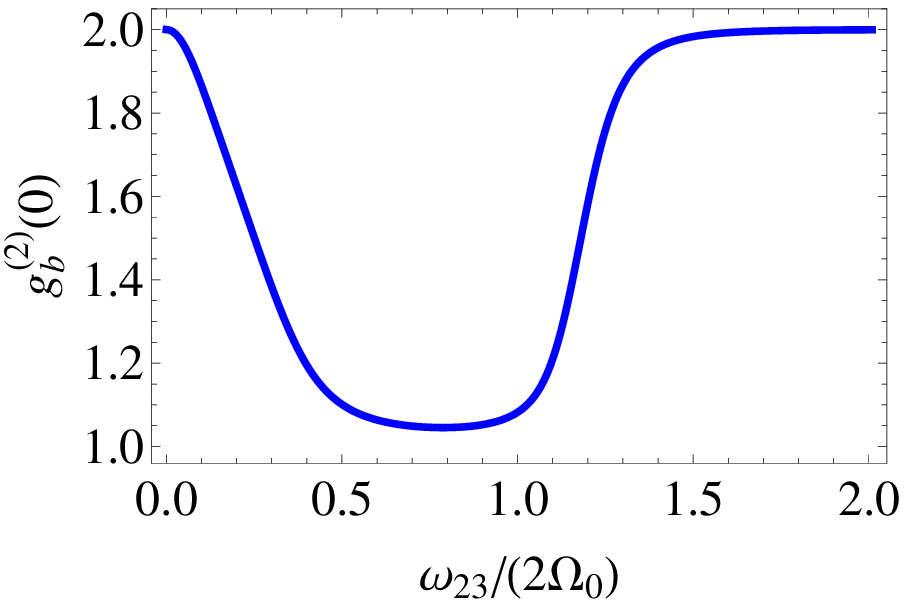}
\begin{picture}(0,0)
\put(-20,80){(a)}
\put(30,80){(b)}
\end{picture}
\caption{\label{fig-2} (a) The mean quanta number of the quantum oscillator $\langle b^{\dagger}b\rangle/\bar n$ and (b) its second-order correlation 
function $g^{(2)}_{b}(0)$ versus $\omega_{23}/(2\Omega_{0})$ for the situation (I). Here $g/\gamma_{2}=4$, $\gamma_{3}/\gamma_{2}=0.1$, 
$\gamma/\gamma_{2}=0$, $\kappa/\gamma_{2}=10^{-3}$, $\omega/\gamma_{2}=50$, $\Omega_{0}/\gamma_{2}=20$ and $\bar n=1$.}
\end{figure}

In order to solve the infinite system of Eq. (\ref{EQM1}), we truncate it at a certain maximum value $n=n_{max}$ so that a further increase of its value, 
i.e. $n_{max}$, does not modify the obtained results. Thus, the steady-state mean quanta's number is expressed as:
\begin{eqnarray}
\langle b^{\dagger} b \rangle = \sum^{n_{max}}_{n=0}nP_{n}^{(0)}, \label{bpb}
\end{eqnarray}
with 
\begin{figure}[t]
\includegraphics[width = 4.23cm]{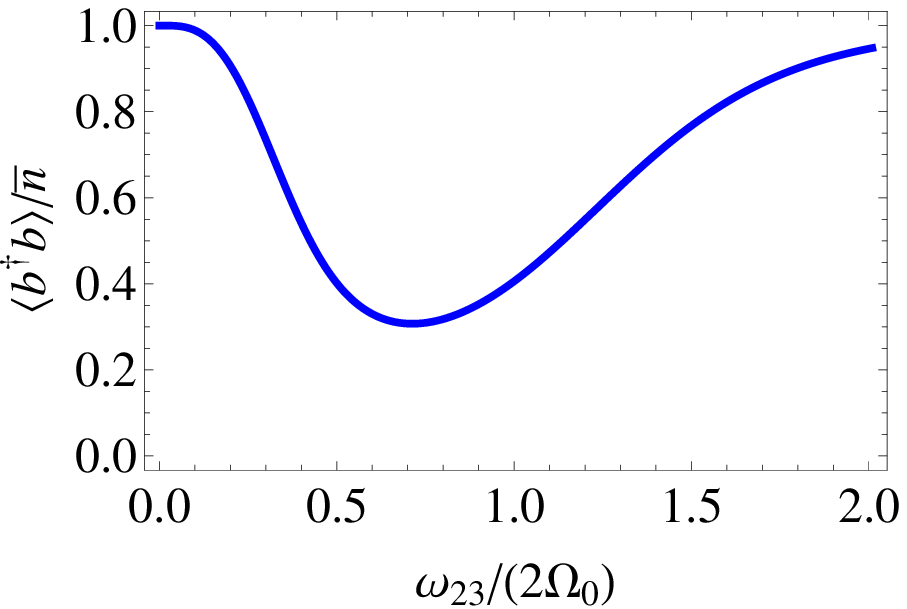}
\includegraphics[width = 4.23cm]{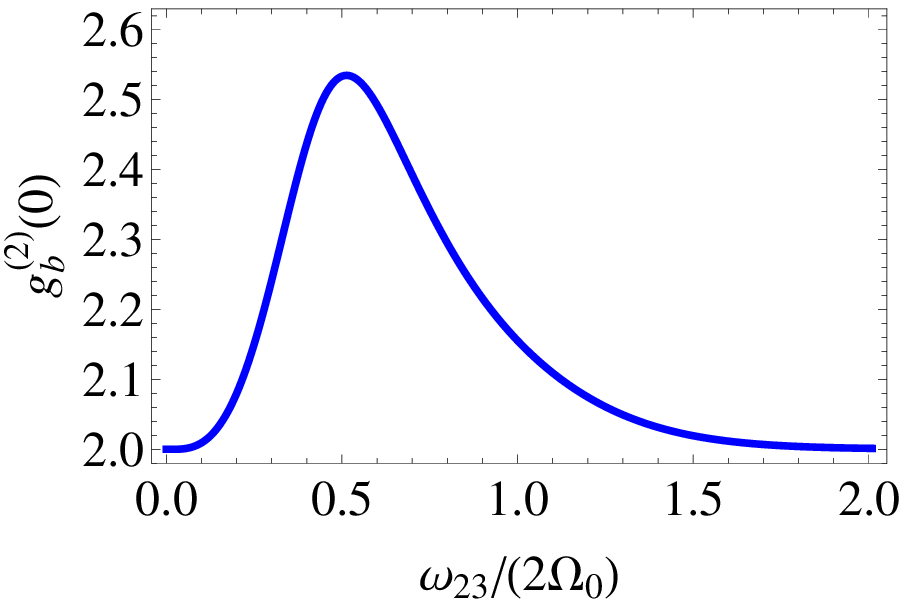}
\begin{picture}(0,0)
\put(-18,70){(a)}
\put(25,70){(b)}
\end{picture}
\caption{\label{fig-3} (a) The scaled mean quanta number of the quantum oscillator $\langle b^{\dagger}b\rangle/\bar n$ and (b) the corresponding 
second-order correlation function $g^{(2)}_{b}(0)$ against the scaled control parameter $\omega_{23}/(2\Omega_{0})$ for the situation (I). 
Here $g/\gamma_{3}=4$, $\gamma_{2}/\gamma_{3}=0.1$, $\gamma/\gamma_{3}=0$, $\kappa/\gamma_{3}=10^{-3}$, $\omega/\gamma_{3}=50$, 
$\Omega_{0}/\gamma_{3}=20$ and $\bar n=15$.}
\end{figure}
\begin{eqnarray}
\sum^{n_{max}}_{n=0}P_{n}^{(0)}=1, \label{nrm}
\end{eqnarray}
while its steady-state second-order correlation function is defined as usual \cite{glauber}, namely,
\begin{eqnarray}
g^{(2)}_{b}(0) &=& \frac{ \langle b^{\dagger}b^{\dagger}bb\rangle }{\langle b^{\dagger} b\rangle^{2} } \nonumber \\
&=&\frac{1}{\langle b^{\dagger}b\rangle ^{2}}\sum^{n_{max}}_{n=0}n(n-1)P_{n}^{(0)}. \label{gg2}
\end{eqnarray}
Respectively, the steady-state mean value of the dressed-state inversion operator, 
$\langle R_{z}\rangle=\langle R_{22}\rangle - \langle R_{33}\rangle$, can be obtained as follows:
\begin{eqnarray}
\langle R_{z}\rangle = \sum^{n_{max}}_{n=0}P^{(2)}_{n}. \label{rzs}
\end{eqnarray}
Figure (\ref{fig-2}) shows the steady-state behaviors of the mean quanta number and its quantum statistics based on Eqs.~(\ref{EQM1}) 
and Exps.~(\ref{bpb},\ref{nrm},\ref{gg2}). The maximum for $\langle b^{\dagger}b\rangle$ occurs around $\bar \delta =0$, i.e., at 
the resonance when the quanta's frequency $\omega$ equals the dressed-state splitting frequency $2\Omega$ due to pumping lasers. 
Importantly here, the quanta's statistics is near Poissonian meaning that we have obtained lasing regimes in our system, see Figs.~\ref{fig-2}(a,b). 
Also, lasing is taking place if $\gamma_{3}/\gamma_{2} \ll 1$. In this case $\langle R_{22} \rangle > \langle R_{33} \rangle$, that is, we have 
dressed-state population inversion and this is the reason for lasing effect, see Fig.~\ref{fig-pI}(a). To avoid any confusion via {\it lasing} 
we mean generation of quantum oscillator's quanta possessing Poissonian statistics, i.e., $g^{(2)}_{b}(0)=1$. Respectively, Figure~(\ref{fig-3}) 
depicts the cooling regimes in this system, under situation (I). This happens when $\gamma_{2}/\gamma_{3} \ll1$ meaning that 
$\langle R_{22} \rangle < \langle R_{33} \rangle$ leading to quanta's absorption processes, see Fig.~\ref{fig-pI}(b). The minimum in the mean quanta 
number followed by an increased second-order correlation function $g^{(2)}_{b}(0)$ occur around $\bar \delta =0$, that is, at resonance 
condition, see Figs.~\ref{fig-3}(a,b).
\begin{figure}[t]
\includegraphics[width = 4.23cm]{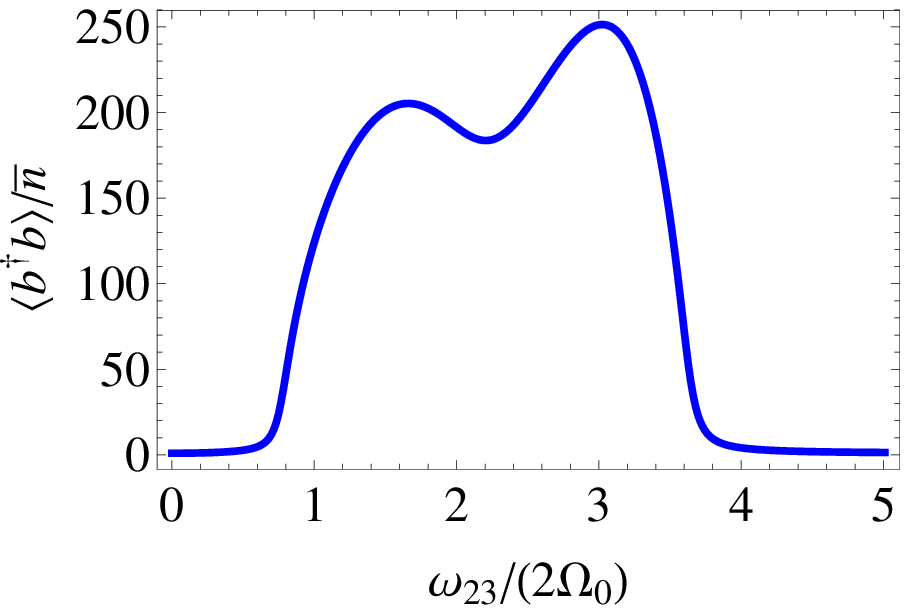}
\includegraphics[width = 4.23cm]{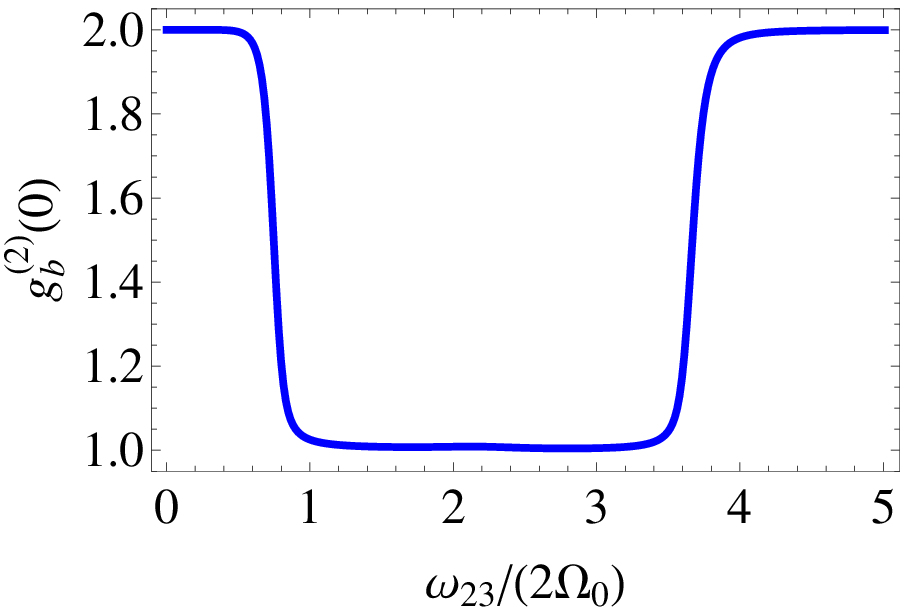}
\begin{picture}(0,0)
\put(-16,78){(a)}
\put(23,78){(b)}
\end{picture}
\caption{\label{fig-4} (a) The mean quanta number of the quantum oscillator $\langle b^{\dagger}b\rangle/\bar n$ and (b) its second-order correlation 
function $g^{(2)}_{b}(0)$ versus $\omega_{23}/(2\Omega_{0})$ for the situation (II) with $\gamma_{3}/\gamma_{2} \ll 1$. All other parameters 
are as in Figure (\ref{fig-2}).}
\end{figure}

\begin{figure}[t]
\includegraphics[width = 4.23cm]{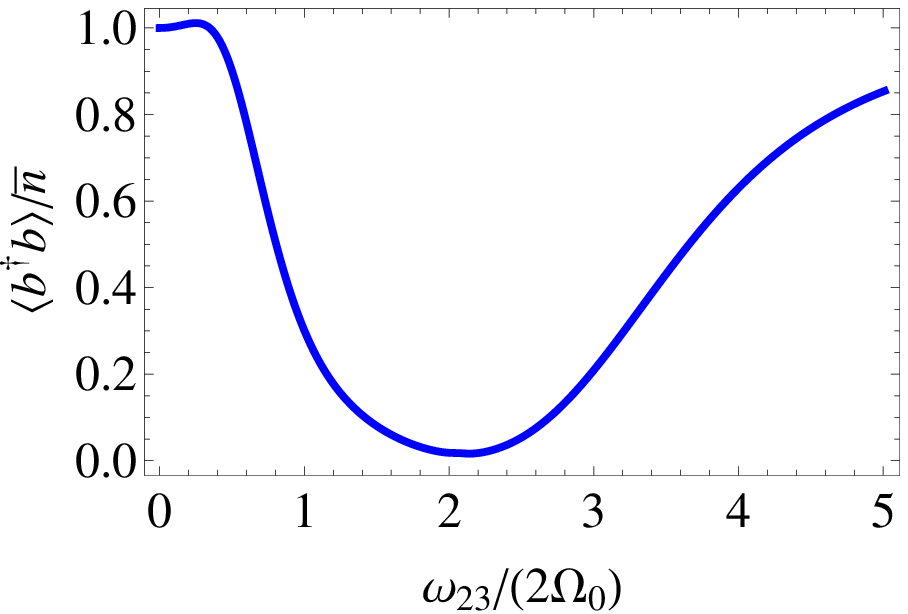}
\includegraphics[width = 4.23cm]{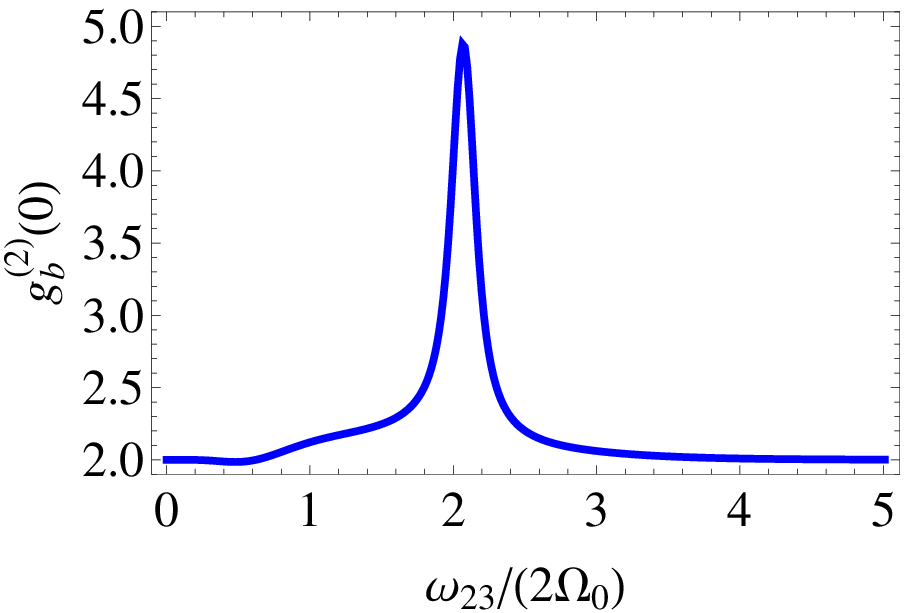}
\begin{picture}(0,0)
\put(-18,80){(a)}
\put(25,80){(b)}
\end{picture}
\caption{\label{fig-5} (a) The scaled mean quanta number of the quantum oscillator $\langle b^{\dagger}b\rangle/\bar n$ and (b) its second-order correlation 
function $g^{(2)}_{b}(0)$ versus $\omega_{23}/(2\Omega_{0})$ for the situation (II) with $\gamma_{2}/\gamma_{3} \ll 1$. 
All other parameters are as in Figure (\ref{fig-3}).}
\end{figure}

Further, for the sake of comparison, we will keep the same parameters and shall investigate the quantum dynamics for the second 
situation, i.e. $(II)$. The respective equations of motion describing the quantum oscillator's dynamics as well as the quantum emitter's 
one are given in Appendix B, i.e., Eqs.~(\ref{aa1}). Particularly, Fig.~\ref{fig-4}(a) shows the mean quanta's number of the quantum 
oscillator in this case, whereas Fig.~\ref{fig-4}(b) depicts the corresponding behavior of the second-order quanta's correlation function 
as a function of $\omega_{23}/(2\Omega_{0})$ when $\gamma_{3}/\gamma_{2} \ll 1$. Remarkably, one can observe a wide plateau 
where the quanta's statistics is Poissonian while its quantum oscillator's mean quanta number vary from small to larger numbers. Thus, 
we have a clear lasing effect in this setup. Compared with the corresponding case, but for the first situation (I), i.e. Fig.~(\ref{fig-2}), 
here, there are generated more quanta of the quantum oscillator followed by a broader lasing regime which is more convenient for 
potential applications, see Fig.~(\ref{fig-4}) and Fig.~(\ref{fig-2}). In this context, if the upper state $|1\rangle$ of the three-level 
emitter has a permanent dipole then it can couple with a single cavity electromagnetic field mode of terahertz frequency, for instance. 
In this case, we have obtained a coherent electromagnetic field source generating terahertz photons. Regarding external applied field 
intensities $I$: For transition wavelengths of the order of $1\mu m$, spontaneous decay rates within the range $10^{9}-10^{10}$Hz, 
and the corresponding THz interval for the Rabi frequencies $\Omega \sim 10^{11} - 10^{12}$Hz, one obtains $I$ within few to several 
$kW/cm^{2}$ which correspond to moderate laser intensities. Respectively, Fig.~\ref{fig-5}(a) emphasizes the cooling regime in the 
examined system, and for the second situation $(II)$, occurring when $\gamma_{2}/\gamma_{3} \ll1$. The second-order correlation 
function increases respectively, see Fig.~\ref{fig-5}(b), demonstrating enhanced phonon-phonon or photon-photon correlations 
depending on the model we have in mind. Compared with Fig.~(\ref{fig-3}) describing same things but for the first situation (I), 
the cooling is significantly enhanced in the second case $(II)$ while keeping identical parameters, see Fig.~(\ref{fig-5}) and Fig.~(\ref{fig-3}). 
The steady-state mean value of dressed-state inversion operator $\langle R_{z}\rangle$, in the lasing regime,  behave differently 
in this case, compare Fig.~\ref{fig-pII}(a) with Fig.~\ref{fig-pI}(a). In the second situation (II), $\langle R_{z}\rangle$ approaches 
zero values, while the mean quanta's number is large, although has a minimum, see Fig.~\ref{fig-4}(a). As we shall explain below, 
these behaviors are due to quantum interference effects. However, cooling occurs for $\langle R_{22} \rangle < \langle R_{33} \rangle$ 
facilitating quanta's absorption processes, see Fig.~\ref{fig-pII}(b). Note that we have carefully checked the convergence of our 
results with respect to various values for $n_{max}$.

Although both situations $(I)$ and $(II)$ show cooling or lasing phenomena, the mechanisms behind them are completely different. 
If $\gamma_{2} \not=\gamma_{3}$ and $\gamma=0$, the first situation $(I)$ resembles a two-level system 
$\{|\Psi_{2}\rangle,|\Psi_{3}\rangle \}$ of frequency $2\Omega$ interacting, respectively, with a quantum oscillator of frequency 
$\omega$, with $2\Omega \approx \omega$, see also \cite{d_las}. The spontaneous decay acts in both directions, i.e. 
$|\Psi_{2}\rangle \leftrightarrow |\Psi_{3}\rangle$, with a corresponding impact on cooling or lasing effects. The cross-correlation 
terms from the Master Equation (\ref{MEQA}) do not influence the quantum dynamics in this case from the simply reason that they 
do not enter at all in the equations of motion (\ref{EQM1}). On the other side, the second situation $(II)$ is close to an equidistant 
three-level system $|\Psi_{2}\rangle \leftrightarrow |\Psi_{1}\rangle \leftrightarrow |\Psi_{3}\rangle$, where each transition being of 
frequency $\Omega$ interacts as well with the quantum oscillator possessing the frequency $\omega$, however, with 
$\Omega \approx \omega$. In this case transitions may take place via single oscillator's quanta processes among the dressed-state 
$|\Psi_{2}\rangle \leftrightarrow |\Psi_{1}\rangle \leftrightarrow |\Psi_{3}\rangle$ or, respectively, involving two-quanta effects among 
the dressed-states $|\Psi_{2}\rangle \leftrightarrow |\Psi_{3}\rangle$. This also means that cross-correlation terms from the Master 
Equation (\ref{MEQA}) do influence the quantum dynamics in this case. This is clearly elucidated also if one inspects the variables 
$\rho^{(i)}, \{i \in 0 \cdots 16\}$, given in the Appendix B, since it contain single or two-quanta processes appearing concomitantly. 
The various decay paths among the dressed-states involved $|\Psi_{2}\rangle \leftrightarrow |\Psi_{1}\rangle \leftrightarrow |\Psi_{3}\rangle$  
lead to quantum interference effects, see also Eq.~(\ref{MEQA}), although the dipole moments corresponding to the two bare transitions 
of the $\Lambda-$type sample are orthogonal to each other. These cross-correlations \cite{agarwal,ficek_book,kiffner} among the dressed-states 
contribute to a more flexible domain for lasing and deeper cooling regimes compared to the situation $(I)$ and for the same parameters 
involved. Thus, one can conclude that quantum interference effects via single- or two-quanta processes distinguish the situation $(II)$, 
described by the Hamiltonian (\ref{DH2}), from the corresponding one characterized by the Hamiltonian (\ref{DH1}), i.e., the case $(I)$. 
This is also the reason that the three-level emitter's population dynamics behave differently as well in these two cases, compare Fig.~(\ref{fig-pI}) 
and Fig.~(\ref{fig-pII}). Notice that when $\omega_{23}/2\Omega_{0}\to 0$ then the quantum emitter lies in the state 
$|\Psi\rangle=\bigl(|3\rangle-|2\rangle\bigr)/\sqrt{2}$, whereas $\langle b^{\dagger}b\rangle/\bar n=1$ and $g^{(2)}_{b}(0)=2$, see 
Figs.~(\ref{fig-2}-\ref{fig-5}), meaning that the quantum oscillator's mode is in a thermal state and no cooling or lasing effects take place, 
respectively. Here, these phenomena occur for $\omega_{23}/2\Omega_{0}\not = 0$, when some population resides on the higher upper 
state $|1 \rangle$, which is distinct from other related schemes based, however, on coherent population trapping effects or electromagnetically 
induced transparency phenomenon \cite{gxl1,gxl2,plenio2,plenio3}. Furthermore, we have observed that there are no cooling effects for both 
cases described here, $(I)$ or $(II)$, if $\gamma_{2}=\gamma_{3}$ while $\gamma=0$. However, the phenomenon it will appear as you 
increases $\gamma$ while keeping $\gamma_{2}=\gamma_{3}$. Finally, the temperatures ranges considered here are within several Kelvins 
for phonon cooling effects to few hundreds of Kelvins for coherent THz photon generation, respectively.

\section{Summary}
Summarizing, we have investigated a laser-pumped three-level $\Lambda-$type system the upper state of which is being coupled with a quantum 
oscillator characterized by a single quantized leaking mode. We have identified two distinct situations leading to cooling or lasing effects of the 
quantum oscillator's degrees of freedom and have described the mechanisms behind them. Particularly, we have demonstrated that the interplay 
between single- or two-quanta processes accompanied by quantum interference effects among the induced emitter's dressed-states are responsible 
for flexible lasing or deeper cooling effects, respectively.  This leads also to mutual influences between the quantum oscillator's dynamics and the 
three-level emitter's quantum dynamics, respectively. The coherent terahertz photons generation is identified as one of the possible application 
resulting from this study.

\acknowledgments
We acknowledge the financial support via grant No. 15.817.02.09F as well as the useful discussions with Victor Ceban, Profirie Bardetski and Corneliu Gherman.

\appendix
\section{The master equation}
\begin{figure}[t]
\includegraphics[width = 4.26cm]{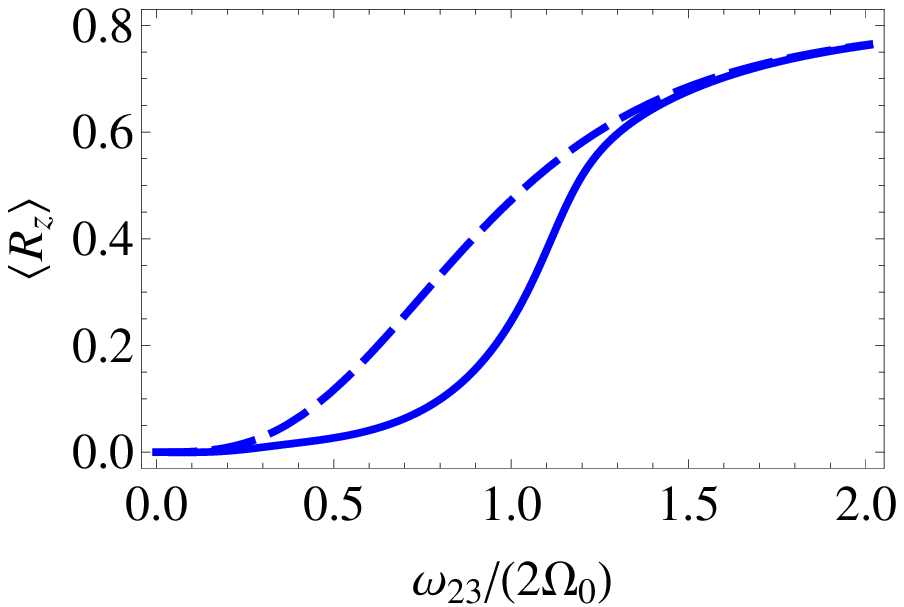}
\hspace{-0.4cm}
\includegraphics[width = 4.35cm]{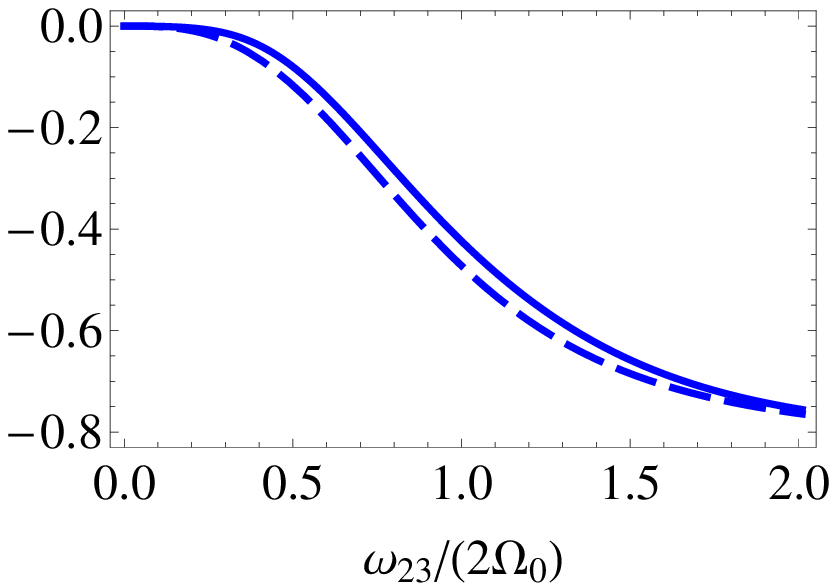}
\begin{picture}(0,0)
\put(-220,70){(a)}
\put(-20,67){(b)}
\end{picture}
\caption{\label{fig-pI} The mean dressed-state inversion operator $\langle R_{z}\rangle = \langle R_{22}\rangle - \langle R_{33}\rangle$ 
as a function of $\omega_{23}/(2\Omega_{0})$ obtained in the steady-state for the first situation (I). (a) $\gamma_{3}/\gamma_{2} \ll 1$ 
whereas (b) $\gamma_{2}/\gamma_{3} \ll 1$. The solid lines are obtained with the full system of equations (\ref{EQM1}), while 
the dashed lines in the absence of the quantum oscillator, i.e. with Exp.~(\ref{ssqp}). All other parameters are as in Fig.~(\ref{fig-2}) and 
Fig.~(\ref{fig-3}), respectively.}
\end{figure}

\begin{figure}[t]
\includegraphics[width = 4.26cm]{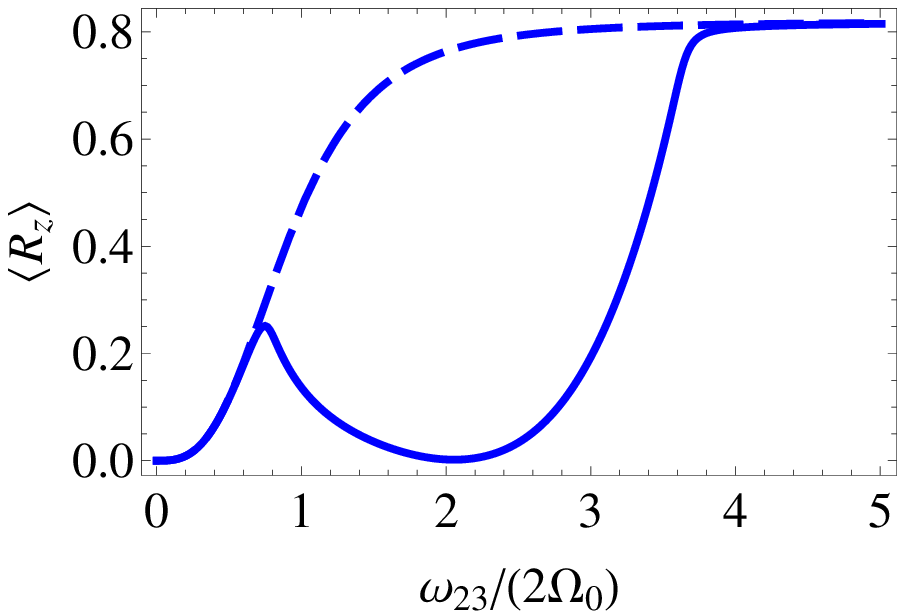}
\hspace{-0.4cm}
\includegraphics[width = 4.35cm]{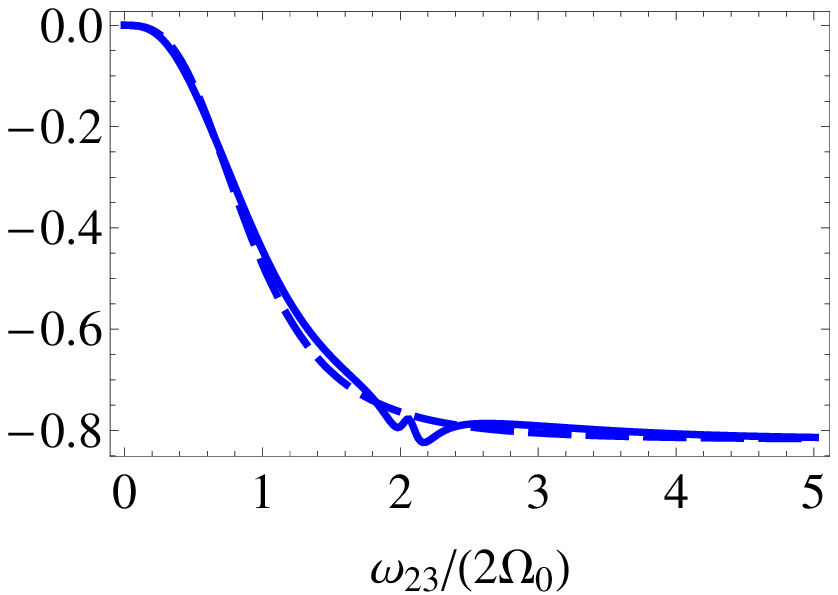}
\begin{picture}(0,0)
\put(-220,70){(a)}
\put(-20,67){(b)}
\end{picture}
\caption{\label{fig-pII} The same as in Fig.~(\ref{fig-pI}) but for the second case (II). The solid lines are obtained with the full system of equations of 
motion (\ref{aa1}), while the dashed lines with Exp.~(\ref{ssqp}). All other parameters are as in Fig.~(\ref{fig-4}) and 
Fig.~(\ref{fig-5}), respectively.}
\end{figure}

Below, one can find the final Master Equation used to obtain the corresponding equations of motion describing the quantum dynamics 
of both the quantum oscillator as well as of the three-level $\Lambda-$type emitter, that is,
\begin{eqnarray}
\dot \rho &+& \frac{i}{\hbar}[H,\rho] = -\gamma_{2}[R^{(+)},R^{(+)}\rho] - \gamma_{3}[R^{(-)},R^{(-)}\rho] \nonumber \\
&-&\frac{\sin^{2}{\theta}}{4}\gamma^{(+)}[R_{12},R_{21}\rho] -\frac{\sin^{2}{\theta}}{4}\gamma^{(-)}[R_{13},R_{31}\rho] 
\nonumber \\
&-& \gamma^{(0)}_{0}\bigl([R_{21},R_{12}\rho] + [R_{31},R_{13}\rho]\bigr) - \Gamma^{(+)}[R_{32},R_{23}\rho] \nonumber \\
&-& \Gamma^{(-)}[R_{23},R_{32}\rho] - \frac{\gamma^{(+)}_{0}}{2}\bigl([R_{12},R_{13}\rho] + [R_{31},R_{21}\rho] \bigr)  \nonumber \\
&-&\frac{\gamma^{(-)}_{0}}{2}\bigl([R_{21},R_{31}\rho] + [R_{13},R_{12}\rho] \bigr) - \frac{\gamma}{4}\cos^{4}{\theta} \nonumber \\
&\times&[\frac{1}{2}(R_{22} +  R_{33})-R_{11},\bigl(\frac{1}{2}(R_{22}+R_{33})- R_{11} \bigr)\rho] \nonumber \\
&-&\frac{\gamma}{8}\cos^{2}{\theta}(1-\sin{\theta})^{2}[R_{12}+R_{31},(R_{21}+R_{13})\rho] \nonumber \\
&-&\frac{\gamma}{8}\cos^{2}{\theta}(1+\sin{\theta})^{2}[R_{21}+R_{13},(R_{12}+R_{31})\rho] \nonumber \\
&-&  \kappa(1+\bar n)[b^{\dagger},b\rho] - \kappa\bar n[b,b^{\dagger}\rho] + H.c.,
 \label{MEQA}
\end{eqnarray}
where $R^{(\pm)}$=$\frac{\sin{2\theta}}{2\sqrt{2}}R_{11} \mp \frac{\cos{\theta}}{2\sqrt{2}}(1 \pm \sin{\theta})R_{22}
\pm \frac{\cos{\theta}}{2\sqrt{2}}(1 \mp \sin{\theta})R_{33}$. The following terms: $[R_{12},R_{13}\rho]$, $[R_{31},R_{21}\rho]$ 
$[R_{21},R_{31}\rho]$ and $[R_{13},R_{12}\rho]$ as well as their Hermitian conjugate parts characterize the cross-damping effects 
or quantum interference phenomena \cite{agarwal,ficek_book,kiffner}. As an exercise, we present the equations of motion for the 
dressed-state populations of the three-level emitter in the absence of the quantum oscillator, that is $g=0$,
\begin{eqnarray}
\langle \dot R_{22}\rangle &=& \gamma^{(+)}_{11}\langle R_{11}\rangle - \gamma^{(+)}_{22}\langle R_{22}\rangle + 
\gamma^{(+)}_{33}\langle R_{33}\rangle, \nonumber \\
\langle \dot R_{33}\rangle &=& \gamma^{(-)}_{11}\langle R_{11}\rangle + \gamma^{(-)}_{33}\langle R_{22}\rangle - 
\gamma^{(-)}_{22}\langle R_{33}\rangle, \nonumber \\
\langle R_{11}\rangle &=& 1- \langle R_{22}\rangle -\langle R_{33}\rangle.
\label{qpop}
\end{eqnarray}
 Here, $\gamma^{(\pm)}_{11}=\gamma^{(\pm)}\sin^{2}{\theta}/2 + \gamma\cos^{2}{\theta}(1 \mp \sin{\theta})^{2}/4$,
$\gamma^{(\pm)}_{22}=2\gamma^{(0)}_{0} + \Gamma^{(\mp)}/2 + \gamma\cos^{2}{\theta}(1 \pm \sin{\theta})^{2}/4$ and 
$\gamma^{(\pm)}_{33}=\gamma^{(\pm)}\cos^{2}{\theta}/4 + \gamma(1 \mp \sin{\theta})^{4}/8$. One can observe that the 
cross-correlation terms from the Master Equation (\ref{MEQA}) do not contribute to population quantum dynamics given by 
Eqs.~(\ref{qpop}). However, their influence will appear in the presence of the quantum oscillator, i.e. when $g \not=0$, and this 
is clearly shown here, compare Fig.~(\ref{fig-pI}) and Fig.~(\ref{fig-pII}).  The steady-state solutions of the above system of 
equations are:
\begin{eqnarray}
\langle R_{22}\rangle &=& \bigl(\gamma^{(+)}_{11}\gamma^{(-)}_{22}+\gamma^{(-)}_{11}\gamma^{(+)}_{33} \bigr)/
\bigl(\gamma^{(+)}_{11}(\gamma^{(-)}_{22}  +  \gamma^{(-)}_{33}) \nonumber \\
&+& \gamma^{(+)}_{22}(\gamma^{(-)}_{11} +  \gamma^{(-)}_{22}) + \gamma^{(+)}_{33}(\gamma^{(-)}_{11} 
-  \gamma^{(-)}_{33}) \bigr), \nonumber \\
\label{ssqp}
\end{eqnarray}
whereas the solution for $\langle R_{33}\rangle$ can be obtained from Exp.~(\ref{ssqp}) via an exchange of upper signs, i.e. 
$(\pm) \to (\mp)$.

Fig.~(\ref{fig-pI}) and Fig.~(\ref{fig-pII}) depict the steady-state values of the dressed-state inversion operator $\langle R_{z}\rangle$ for the 
both cases studied here, $(I)$ and $(II)$, and in the presence of the quantum oscillator (solid lines) as well as in its absence (dashed curves), 
respectively. One can observe that there is a clear difference between the cases with $g=0$ and $g\not=0$ in the lasing regimes, 
compare Fig.~\ref{fig-pI}(a) and Fig.~\ref{fig-pII}(a). As it was described above, this distinction is due to cross-correlation terms or quantum 
interference effects arising in the second case (II). Correspondingly, in the cooling regimes the quantum oscillator's influence on the steady-state 
mean value of the qubit inversion operator is not quite significant, although still visible.

\section{The equations of motion when $\omega \approx \Omega$, i.e., for the case (II)}
Here, we shall present the equations of motion for the second situation $(II)$ obtained with the help of the Master Equation (\ref{MEQA}), 
that is,
\begin{eqnarray}
\dot P^{(0)}_{n} &=& i\tilde g(P^{(3)}_{n} - P^{(5)}_{n}-P^{(9)}_{n} + P^{(7)}_{n}) - 2\kappa\bar n\bigl((n+1)P^{(0)}_{n} \nonumber \\
&-& nP^{(0)}_{n-1}\bigr) - 2\kappa(1+\bar n)\bigl(nP^{(0)}_{n} - (n+1)P^{(0)}_{n+1}\bigr), \nonumber \\
\dot P^{(1)}_{n} &=& i\tilde g(P^{(7)}_{n} - P^{(9)}_{n}) - 2\kappa\bar n\bigl((n+1)P^{(1)}_{n} - nP^{(1)}_{n-1}\bigr) \nonumber \\
&-&2\kappa(1+\bar n)\bigl(nP^{(1)}_{n} - (n+1)P^{(1)}_{n+1}\bigr) + \tilde \gamma^{(1)}_{0}P^{(0)}_{n} \nonumber \\
&-& \tilde \gamma^{(1)}_{1}P^{(1)}_{n} - \tilde \gamma^{(1)}_{2}P^{(2)}_{n}, \nonumber 
\end{eqnarray}
\begin{eqnarray}
\dot P^{(2)}_{n} &=& -i\tilde g(P^{(9)}_{n} + P^{(7)}_{n}) - 2\kappa\bar n\bigl((n+1)P^{(2)}_{n}- nP^{(2)}_{n-1}\bigr)\nonumber \\
&-& 2\kappa(1+\bar n)\bigl(nP^{(2)}_{n} - (n+1)P^{(2)}_{n+1}\bigr) + \tilde \gamma^{(2)}_{0}P^{(0)}_{n} \nonumber \\
&+& \tilde \gamma^{(2)}_{1}P^{(1)}_{n} - \tilde \gamma^{(2)}_{2}P^{(2)}_{n}, \nonumber \\
\dot P^{(3)}_{n} &=& i\tilde \delta P^{(4)}_{n}  - \tilde \gamma^{(3)}_{3}P^{(3)}_{n} + \tilde \gamma^{(3)}_{7}P^{(7)}_{n} \nonumber \\
&+& i\tilde g(n(2P^{(0)}_{n}- P^{(1)}_{n-1}-P^{(2)}_{n-1})-(2n+1)P^{(1)}_{n}) \nonumber \\
&-&\kappa(1+\bar n)\bigl((2n-1)P^{(3)}_{n}-2 (n+1)P^{(3)}_{n+1} \nonumber \\
&+& 2P^{(9)}_{n}\bigr) - \kappa\bar n\bigl((2n+1)P^{(3)}_{n} - 2nP^{(3)}_{n-1}\bigr), \nonumber \\
\dot P^{(4)}_{n} &=& i\tilde \delta P^{(3)}_{n} - i\tilde g P^{(12)}_{n} - \kappa(1+\bar n)\bigl((2n-1)P^{(4)}_{n} \nonumber \\
&+&2P^{(10)}_{n} -2 (n+1)P^{(4)}_{n+1}\bigr) - \kappa\bar n\bigl((2n+1)P^{(4)}_{n} \nonumber \\
&-& 2nP^{(4)}_{n-1}\bigr)- \tilde \gamma^{(4)}_{4}P^{(4)}_{n} + \tilde \gamma^{(4)}_{8}P^{(8)}_{n}, \nonumber \\
\dot P^{(5)}_{n}&=& i\tilde \delta P^{(6)}_{n} + i\tilde g\bigl(P^{(11)}_{n}+(n+1)( P^{(1)}_{n+1} - P^{(2)}_{n+1}) \nonumber \\
&-& 2(n+1)( P^{(0)}_{n}-P^{(1)}_{n})\bigr) \nonumber \\
&-& \kappa(1+\bar n)\bigl((2n+1)P^{(5)}_{n} - 2 (n+1)P^{(5)}_{n+1}\bigr) \nonumber \\
&-& \kappa\bar n\bigl((2n+3)P^{(5)}_{n} - 2nP^{(5)}_{n-1} - 2P^{(7)}_{n}\bigr)  \nonumber \\
&-&\tilde\gamma^{(5)}_{5}P^{(5)}_{n} + \tilde\gamma^{(5)}_{9}P^{(9)}_{n}, \nonumber \\
\dot P^{(6)}_{n} &=& i\tilde \delta P^{(5)}_{n} + i\tilde g P^{(12)}_{n}-  \kappa\bar n\bigl((2n+3)P^{(6)}_{n} - 2nP^{(6)}_{n-1} \nonumber \\
&-& 2P^{(8)}_{n}\bigr) - \kappa(1+\bar n)\bigl((2n+1)P^{(6)}_{n} - 2(n+1) \nonumber \\
&\times&P^{(6)}_{n+1}\bigr) - \tilde \gamma^{(6)}_{6}P^{(6)}_{n} + \tilde \gamma^{(6)}_{10}P^{(10)}_{n}, \nonumber \\
\dot P^{(7)}_{n}&=& i\tilde \delta P^{(8)}_{n} + i\tilde g\bigl(P^{(13)}_{n}+n( P^{(1)}_{n} - P^{(2)}_{n}) - 2n( P^{(0)}_{n-1} \nonumber \\
&-& P^{(1)}_{n-1})\bigr) -\kappa\bar n\bigl((2n+1)P^{(7)}_{n} - 2nP^{(7)}_{n-1}\bigr) \nonumber \\
&-& \kappa(1+\bar n)\bigl((2n-1)P^{(7)}_{n} - 2(n+1)P^{(7)}_{n+1} \nonumber \\
&+&2P^{(5)}_{n} \bigr) + \tilde\gamma^{(7)}_{3}P^{(3)}_{n} - \tilde\gamma^{(7)}_{7}P^{(7)}_{n}, \nonumber \\
\dot P^{(8)}_{n}&=& i\tilde \delta P^{(7)}_{n}+ i\tilde g P^{(14)}_{n} - \kappa\bar n\bigl((2n+1)P^{(8)}_{n} - 2nP^{(8)}_{n-1} \bigr) \nonumber \\
&-& \kappa(1+\bar n)\bigl((2n-1)P^{(8)}_{n} - 2(n+1)P^{(8)}_{n+1} \nonumber \\
&+&2P^{(6)}_{n}\bigr) + \tilde \gamma^{(8)}_{4}P^{(4)}_{n} - \tilde \gamma^{(8)}_{8}P^{(8)}_{n}, \nonumber \\
\dot P^{(9)}_{n}&=& i\tilde \delta P^{(10)}_{n} + i\tilde g\bigl(2(n+1)( P^{(0)}_{n+1} - P^{(1)}_{n+1}) - (n+1) \nonumber \\
&\times& ( P^{(1)}_{n} + P^{(2)}_{n}) - P^{(15)}_{n}\bigr) - \kappa(1+\bar n)\bigl((2n+1)P^{(9)}_{n} \nonumber \\
&-& 2(n+1)P^{(9)}_{n+1} \bigr) - \kappa\bar n\bigl((2n+3)P^{(9)}_{n} - 2nP^{(9)}_{n-1} \nonumber \\
&-& 2P^{(3)}_{n} \bigr) + \tilde\gamma^{(9)}_{5}P^{(5)}_{n} - \tilde\gamma^{(9)}_{9}P^{(9)}_{n}, \nonumber \\
\dot P^{(10)}_{n}&=& i\tilde \delta P^{(9)}_{n}- i\tilde g P^{(16)}_{n} - \kappa\bar n\bigl((2n+3)P^{(10)}_{n} - 2nP^{(10)}_{n-1} \nonumber \\
&-& 2P^{(4)}_{n}\bigr) - \kappa(1+\bar n)\bigl((2n+1)P^{(10)}_{n} - 2(n+1) \nonumber \\
&\times& P^{(10)}_{n+1} \bigr) + \tilde \gamma^{(10)}_{6}P^{(6)}_{n} - \tilde \gamma^{(10)}_{10}P^{(10)}_{n}, \nonumber \\
\dot P^{(11)}_{n}&=& 2i\tilde \delta P^{(12)}_{n} + i\tilde g\bigl(n P^{(5)}_{n} - (n+1) P^{(3)}_{n}\bigr) - 2\kappa(1+\bar n)\nonumber \\
&\times& \bigl(nP^{(11)}_{n}- (n+1)P^{(11)}_{n+1} + P^{(15)}_{n}\bigr) - 2\kappa\bar n\bigl((n+1) \nonumber \\
&\times& P^{(11)}_{n} - nP^{(11)}_{n-1} - P^{(13)}_{n}\bigr) - \tilde\gamma^{(11)}_{11}P^{(11)}_{n}, \nonumber \\
\dot P^{(12)}_{n}&=& 2i\tilde \delta P^{(11)}_{n} + i\tilde g\bigl(n P^{(6)}_{n} - (n+1) P^{(4)}_{n}\bigr) - 2\kappa(1+\bar n)\nonumber \\
&\times& \bigl(nP^{(12)}_{n}- (n+1)P^{(12)}_{n+1} + P^{(16)}_{n}\bigr) - 2\kappa\bar n\bigl((n+1) \nonumber \\
&\times& P^{(12)}_{n} - nP^{(12)}_{n-1} - P^{(14)}_{n}\bigr) - \tilde\gamma^{(12)}_{12}P^{(12)}_{n}, \nonumber 
\end{eqnarray}
\begin{eqnarray}
\dot P^{(13)}_{n}&=&2i\tilde \delta P^{(14)}_{n} + i\tilde g\bigl((n-1)P^{(7)}_{n} - nP^{(3)}_{n-1}\bigr)-2\kappa(1+\bar n)\nonumber \\
&\times& \bigl((n-1)P^{(13)}_{n} - (n+1)P^{(13)}_{n+1}+2P^{(11)}_{n}\bigr)-2\kappa n\bar n \nonumber \\
&\times& \bigl(P^{(13)}_{n} - P^{(13)}_{n-1}\bigr) - \tilde\gamma^{(13)}_{13}P^{(13)}_{n}, \nonumber \\
\dot P^{(14)}_{n}&=&2i\tilde \delta P^{(13)}_{n}+ i\tilde g\bigl((n-1)P^{(8)}_{n}- nP^{(4)}_{n-1}\bigr)-2\kappa(1+\bar n)\nonumber \\
&\times& \bigl((n-1)P^{(14)}_{n} - (n+1)P^{(14)}_{n+1}+2P^{(12)}_{n}\bigr)-2\kappa n\bar n \nonumber \\
&\times& \bigl(P^{(14)}_{n} - P^{(14)}_{n-1}\bigr) - \tilde\gamma^{(14)}_{14}P^{(14)}_{n}, \nonumber \\
\dot P^{(15)}_{n}&=&2i\tilde \delta P^{(16)}_{n}+ i\tilde g\bigl((n+1)P^{(5)}_{n}-(n+2)P^{(9)}_{n}\bigr) \nonumber \\
&-&2\kappa(1+\bar n)(1+n)\bigl(P^{(15)}_{n} - P^{(15)}_{n+1}\bigr)-2\kappa \bar n \nonumber \\
&\times& \bigl((n+2)P^{(15)}_{n} -n P^{(15)}_{n-1}- 2P^{(11)}_{n}\bigr) - \tilde\gamma^{(15)}_{15}P^{(15)}_{n}, \nonumber \\
\dot P^{(16)}_{n}&=&2i\tilde \delta P^{(15)}_{n}+ i\tilde g\bigl((n+1)P^{(6)}_{n+1}-(n+2)P^{(10)}_{n}\bigr) \nonumber \\
&-&2\kappa(1+\bar n)(1+n)\bigl(P^{(16)}_{n} - P^{(16)}_{n+1}\bigr)-2\kappa \bar n \nonumber \\
&\times& \bigl((n+2)P^{(16)}_{n} -n P^{(16)}_{n-1}- 2P^{(12)}_{n}\bigr) - \tilde\gamma^{(16)}_{16}P^{(16)}_{n}. \nonumber \\
\label{aa1}
\end{eqnarray}
Here $\tilde \gamma^{(1)}_{0}=\gamma^{(1)}_{0}$, $\tilde \gamma^{(1)}_{1}=\gamma^{(1)}_{1}$, 
$\tilde \gamma^{(1)}_{2}=\gamma\sin{\theta}\cos^{2}{\theta}/2$, $\tilde \gamma^{(2)}_{0}=\gamma^{(2)}_{0}$,
$\tilde \gamma^{(2)}_{1}=-\gamma^{(2)}_{1}$, $\tilde \gamma^{(2)}_{2}=\gamma^{(2)}_{2}$, 
$\tilde \gamma^{(3)}_{3}=\gamma_{2}\cos^{2}{\theta}(1+3\sin{\theta})^{2}/8$ + $\gamma_{3}\cos^{2}{\theta}(1-3\sin{\theta})^{2}/8$ 
+ $(\gamma^{(+)}+\gamma^{(-)})\sin^{2}{\theta}/4 + \gamma^{(0)}_{0}+\Gamma^{(-)}$ + $9\gamma\cos^{4}{\theta}/16$ + 
$\gamma\cos^{2}{\theta}\bigl((1+\sin{\theta})^{2}+ (1-\sin{\theta})^{2}/2\bigr)/4$, $\tilde \gamma^{(3)}_{7}=\gamma^{(+)}_{0}
+\gamma\cos^{2}{\theta}(1-\sin{\theta})^{2}/4$, $\tilde \gamma^{(4)}_{4}=\tilde \gamma^{(3)}_{3}$, $\tilde \gamma^{(4)}_{8} =\tilde \gamma^{(3)}_{7}$,
$\tilde \gamma^{(5)}_{5}=\gamma_{2}\cos^{2}{\theta}(1-3\sin{\theta})^{2}/8$ + $\gamma_{3}\cos^{2}{\theta}(1+3\sin{\theta})^{2}/8$ 
+ $(\gamma^{(+)}+\gamma^{(-)})\sin^{2}{\theta}/4 + \gamma^{(0)}_{0}+\Gamma^{(+)}$ + $9\gamma\cos^{4}{\theta}/16$ + 
$\gamma\cos^{2}{\theta}\bigl((1-\sin{\theta})^{2}+ (1+\sin{\theta})^{2}/2\bigr)/4$, $\tilde \gamma^{(5)}_{9}=\gamma^{(-)}_{0}
+\gamma\cos^{2}{\theta}(1+\sin{\theta})^{2}/4$, $\tilde \gamma^{(6)}_{6}=\tilde \gamma^{(5)}_{5}$,  $\tilde \gamma^{(6)}_{10} =\tilde \gamma^{(5)}_{9}$,
$\tilde \gamma^{(7)}_{7}=\tilde \gamma^{(6)}_{6}$, $\tilde \gamma^{(7)}_{3} =\tilde \gamma^{(6)}_{10}$,
$\tilde \gamma^{(8)}_{8}=\tilde \gamma^{(7)}_{7}$, $\tilde \gamma^{(8)}_{4} =\tilde \gamma^{(7)}_{3}$,
$\tilde \gamma^{(9)}_{5}=\gamma^{(+)}_{0} + \gamma\cos^{2}{\theta}(1 - \sin{\theta})^{2}/4$, 
$\tilde \gamma^{(9)}_{9}=\tilde \gamma^{(3)}_{3}=\tilde \gamma^{(10)}_{10}$, $\tilde \gamma^{(10)}_{6} =\tilde \gamma^{(9)}_{5}$,
$\tilde \gamma^{(11)}_{11}=(\gamma_{2} + \gamma_{3})\cos^{2}{\theta}/2 + 2\gamma^{(0)}_{0} + \Gamma^{(-)} + \Gamma^{(+)} 
+ \gamma\cos^{2}{\theta}(1+\sin^{2}{\theta})/4$, and $\tilde \gamma^{(11)}_{11}=\tilde \gamma^{(12)}_{12}=\tilde \gamma^{(13)}_{13}
=\tilde \gamma^{(14)}_{14}=\tilde \gamma^{(15)}_{15}=\tilde \gamma^{(16)}_{16}$. 

The system of equations (\ref{aa1}) can be obtained if one first get the equations of motion for the variables: 
$\rho^{(0)}=\rho_{11}+\rho_{22}+\rho_{33}$, $\rho^{(1)}=\rho_{22}+\rho_{33}$, $\rho^{(2)}=\rho_{22}-\rho_{33}$, 
$\rho^{(3)}=b^{\dagger}\rho_{21}-\rho_{12}b$, $\rho^{(4)}=b^{\dagger}\rho_{21}+ \rho_{12}b$, 
$\rho^{(5)}=\rho_{13}b^{\dagger} - b\rho_{31}$, $\rho^{(6)}=\rho_{13}b^{\dagger} + b\rho_{31}$, 
$\rho^{(7)}=b^{\dagger}\rho_{13}-\rho_{31}b$, $\rho^{(8)}=b^{\dagger}\rho_{13} + \rho_{31}b$,
$\rho^{(9)}=\rho_{21}b^{\dagger} - b\rho_{12}$, $\rho^{(10)}=\rho_{21}b^{\dagger} + b\rho_{12}$, 
$\rho^{(11)}=b^{\dagger}\rho_{23}b^{\dagger} + b\rho_{32}b$, $\rho^{(12)}=b^{\dagger}\rho_{23}b^{\dagger} - b\rho_{32}b$, 
$\rho^{(13)}=b^{\dagger 2}\rho_{23} + \rho_{32}b^{2}$, $\rho^{(14)}=b^{\dagger 2}\rho_{23} - \rho_{32}b^{2}$,
$\rho^{(15)}=\rho_{23}b^{\dagger 2} + b^{2}\rho_{32}$, $\rho^{(16)}=\rho_{23}b^{\dagger 2} - b^{2}\rho_{32}$, 
using the Master Equation (\ref{MEQA}) and then projecting them on the Fock states $|n\rangle$, i.e., 
$P^{(i)}_{n}=\langle n|\rho^{(i)}|n\rangle$, $\{i \in 0 \cdots 16\}$, and $n \in \{0, \infty\}$. 
Together with Exps.~(\ref{bpb},\ref{nrm},\ref{gg2},\ref{rzs},\ref{ssqp}) one can obtain the interested quantities like the mean quanta's number of 
the quantum oscillator or its quantum statistics described by the second-order correlation function as well as qubit's populations, see Figs.~(\ref{fig-4}), 
(\ref{fig-5}), (\ref{fig-pI}) and (\ref{fig-pII}). 


\begin{thebibliography}{55}
\bibitem{rew_art1} D. Leibfried, R. Blatt, C. Monroe, and D. Wineland, Quantum dynamics of single trapped ions, 
Rev. Mod. Phys. {\bf 75}, 281 (2003).
\bibitem{n_las} M. Khajavikhan, A. Simic, M. Katz, J. H. Lee, B. Slutsky, A. Mizrahi, V. Lomakin, and Y. Fainman,
Thresholdless nanoscale coaxial lasers, Nature {\bf 482}, 204 (2012). 
\bibitem{d_las} J. Zakrzewski, M. Lewenstein, and Th. W.  Mossberg, Theory of dressed-state lasers. I. Effective Hamiltonians and stability 
properties, Phys. Rev. A {\bf 44}, 7717 (1991).
\bibitem{las_c} I. Wilson-Rae, P. Zoller, and A. Imamoglu, Laser Cooling of a Nanomechanical Resonator Mode to its Quantum Ground State,
Phys. Rev. Lett. {\bf 92}, 075507 (2004).
\bibitem{fon_l} J. Kabuss, A. Carmele, T. Brandes, and A. Knorr, Optically Driven Quantum Dots as Source of Coherent Cavity Phonons:
A Proposal for a Phonon Laser Scheme, Phys. Rev. Lett. {\bf 109}, 054301 (2012).
\bibitem{qt} M. Fleischhauer, A. Imamoglu, and J. P. Marangos, Electromagnetically induced transparency: Optics in coherent media, 
Rev. Mod. Phys. {\bf 77}, 633 (2005).
\bibitem{qt1} H. J. Kimble, The quantum internet,  Nature  {\bf 453}, 1023 (2008). 
\bibitem{qt2} D. D. Awschalom, R. Hanson, J. Wrachtrup, and B. B. Zhou, Quantum technologies with optically interfaced
solid-state spins, Nature Photonics {\bf 12}, 516 (2018).
\bibitem{eit_theor} G. Morigi, J. Eschner, and C. H. Keitel, Ground State Laser Cooling Using Electromagnetically Induced 
Transparency, Phys. Rev. Lett. {\bf 85}, 4458 (2000).
\bibitem{eit_cool} C. F. Roos, D. Leibfried, A.Mundt,  F. Schmidt-Kaler, J. Eschner, and R. Blatt, 
Experimental Demonstration of Ground State Laser Cooling with Electromagnetically Induced Transparency, 
Phys. Rev. Lett. {\bf 85}, 5547 (2000).
\bibitem{ek} J. Evers, and C. H. Keitel, Double-EIT ground-state laser cooling without blue-sideband heating, 
Europhys. Lett. {\bf 68}, 370 (2004).
\bibitem{plenio1} J. Cerrillo, A. Retzker, and M. B. Plenio, Fast and Robust Laser Cooling of Trapped Systems, 
Phys. Rev. Lett. {\bf 104}, 043003 (2010).
\bibitem{rew_art2} Y. Greenberg, Y. Pashkin, and E. Il’ichev, Nanomechanical resonators, Phys. Usp. {\bf 55}, 382 (2012).
\bibitem{rew_art3} M. Aspelmeyer, T. J. Kippenberg, and F. Marquardt, Cavity optomechanics, Rev. Mod. Phys. {\bf 86}, 1391 (2014).
\bibitem{xe} K. Xia, and J. Evers, Ground State Cooling of a Nanomechanical Resonator in the Nonresolved Regime via Quantum Interference, 
Phys. Rev. Lett. {\bf 103}, 227203 (2009).
\bibitem{phon_las} K. Vahala, M. Herrmann, S. Kn\"{u}nz, V. Batteiger, G. Saathoff, T. W. H\"{a}nsch, and Th. Udem, A phonon laser, 
Nature Physics {\bf 5}, 682 (2009).
\bibitem{cool_exp1} C. Sch\"{a}fermeier, H. Kerdoncuff, U. B. Hoff, H. Fu, A. Huck, J. Bilek, G. I. Harris, W. P. Bowen, T. Gehring, and U. L. 
Andersen, Quantum enhanced feedback cooling of a mechanical oscillator using nonclassical light, Nature Communications {\bf 7:13628} (2016).
\bibitem{cool_exp2} J. B. Clark, F. Lecocq, R. W. Simmonds, J. Aumentado, and J. D. Teufel, Sideband cooling beyond the quantum backaction
limit with squeezed light, Nature {\bf 541}, 191 (2017).
\bibitem{qwell} G. B. Serapiglia, E. Paspalakis, C. Sirtori, K. L. Vodopyanov, and C. C. Phillips, Laser-Induced Quantum Coherence in a Semiconductor 
Quantum Well, Phys. Rev. Lett. {\bf 84}, 1019 (2000).
\bibitem{qdots} A. Beveratos, I. Abram, J.-M. Gerard, and I. Robert-Philip, Quantum optics with quantum dots:
Towards semiconductor sources of quantum light for quantum information processing, Eur. Phys. J. D {\bf 68}, 377 (2014).
\bibitem{victor} V. Ceban, Phase-dependent quantum interferences with three-level artificial atoms, Romanian Journal of Physics {\bf 62}, 207 (2017).
\bibitem{gxl1} J.-p. Zhu, and G.-x. Li, Ground-state cooling of a nanomechanical resonator with a triple quantum dot via quantum interference, 
Phys. Rev. A {\bf 86}, 053828 (2012).
\bibitem{gxl2} G.-x. Li, and J.-p. Zhu, Ground-state cooling of a mechanical resonator coupled to two coupled quantum dots,  
J. Phys. B: At. Mol. Opt. Phys. {\bf 44}, 195502 (2011).
\bibitem{plenio2} H.-K. Lau, and M. B. Plenio, Laser cooling of a high-temperature oscillator by a three-level system, 
Phys. Rev. B {\bf 94}, 054305 (2016).
\bibitem{plenio3} M. Abdi, and M. B. Plenio, Quantum Effects in a Mechanically Modulated Single-Photon Emitter, 
Phys. Rev. Lett. {\bf 122}, 023602 (2019).
\bibitem{vpm} V. Ceban, P. Longo, and M. A. Macovei, Fast phonon dynamics of a nanomechanical oscillator due to cooperative effects,
Phys. Rev. A {\bf 95}, 023806 (2017).
\bibitem{thz1} G. Ducournau, Terahertz science: Silicon photonics targets terahertz region, Nature Photonics {\bf 12}, 574 (2018).
\bibitem{thz2} T. Harter, S. Muehlbrandt, S. Ummethala, A. Schmid, S. Nellen, L. Hahn, W. Freude, and C. Koos, 
Silicon–plasmonic integrated circuits for terahertz signal generation and coherent detection, Nature Photonics {\bf 12}, 625 (2018).
\bibitem{thz3} Sh. Du, K. Yoshida, Ya Zhang, I. Hamada, and K. Hirakawa, Terahertz dynamics of electron–vibron coupling in single molecules with 
tunable electrostatic potential, Nature Photonics {\bf 12}, 608 (2012).
\bibitem{tera1} O. V. Kibis, G. Ya. Slepyan, S. A. Maksimenko, and A. Hoffmann, Matter Coupling to Strong Electromagnetic Fields in Two-Level 
Quantum Systems with Broken Inversion Symmetry, Phys. Rev. Lett. {\bf 102}, 023601 (2009).
\bibitem{tera2} F. Oster, C. H. Keitel, and M. Macovei, Generation of correlated photon pairs in different frequency ranges,
Phys. Rev. A {\bf 85}, 063814 (2012).
\bibitem{terra} S. De Liberato, C. Ciuti and Ch. C. Phillips, Terahertz lasing from intersubband polariton-polariton scattering 
in asymmetric quantum wells, Phys. Rew. B {\bf 87}, 241304(R) (2013).
\bibitem{tera3} M. Miri, F. Zamani, and H. Alipoor, Two tunneling-coupled two-level systems with broken inversion symmetry: 
tuning the terahertz emission, Jr. Opt. Soc. Am. B {\bf 33}, 1873 (2016). 
\bibitem{tera4} I. Yu. Chestnov, V. A. Shahnazaryan, A. P. Alodjants, and I. A. Shelykh, Terahertz Lasing in Ensemble of Asymmetric Quantum Dots, 
ACS Photonics {\bf 4(11)}, 2726 (2017).
\bibitem{mmk} M. Macovei, M. Mishra, and C. H. Keitel, Population inversion in two-level systems possessing permanent dipoles,
Phys. Rev. A {\bf 92}, 013846 (2015).
\bibitem{gagik} G. Yu. Kryuchkyan, V. Shahnazaryan, O. V. Kibis, and I. A. Shelykh, Resonance fluorescence from an asymmetric quantum dot 
dressed by a bichromatic electromagnetic field, Phys. Rev. A {\bf 95}, 013834 (2017).
\bibitem{paspalakis} M. A. Anton, S. Maede-Razavi, F. Carreno, I. Thanopulos, and E. Paspalakis, Optical and microwave control of resonance 
fluorescence and squeezing spectra in a polar molecule, Phys. Rev. A {\bf 96}, 063812  (2017).
\bibitem{qpd} D. Hiluf, and Y. Dubi, Phonon as environmental disturbance in three level system, arXiv:1803.08327v1.
\bibitem{gong} F. Zhou, Y. Niu, and S. Gong, Electromagnetically induced transparency in a three-level lambda system
with permanent dipole moments, J. Chem. Phys. {\bf 131}, 034105 (2009).
\bibitem{qw} S. Kocinac, Z. Ikonic, and V. Milanovic, Second Harmonic Generation at the Quantum-Interference Induced Transparency
in Semiconductor Quantum Wells: The Influence of Permanent Dipole Moments, IEEE Jr. Quant. Electr. {\bf 37}(7), 873 (2001).
\bibitem{agarwal_2014} G. S. Agarwal, Quantum Optics (Cambridge University Press, 2014).
\bibitem{gardiner} C. W. Gardiner and P. Zoller, Quantum Noise (Springer, Berlin, 2004).
\bibitem{wals} D. F. Walls and G. J. Milburn, Quantum Optics (Springer, Berlin, 2008).
\bibitem{quang} T. Quang and H. Freedhoff, Atomic population inversion and enhancement of resonance fluorescence in a cavity, 
Phys. Rev. A {\bf 47}, 2285 (1993).
\bibitem{glauber} R. J. Glauber, The Quantum Theory of Optical Coherence, Phys. Rev. {\bf 130}, 2529 (1963).
\bibitem{agarwal} G. S. Agarwal, Quantum Statistical Theories of Spontaneous Emission and their Relation to other Approaches 
(Springer, Berlin, 1974).
\bibitem{ficek_book} Z. Ficek and S. Swain, Quantum Interference and Coherence: Theory and Experiments (Springer, Berlin, 2005).
\bibitem{kiffner} M. Kiffner, M. Macovei, J. Evers, and C. H. Keitel, Vacuum induced processes in multilevel atoms, Prog. Opt. {\bf 55}, 85 (2010).
\end{thebibliography}
\end{document}